\documentclass{emulateapj}

\usepackage{amssymb,gensymb,graphics,textcomp}
\usepackage[section] {placeins}

\newcommand{\kms}{\,\mbox{km s$^{-1}$}}
\newcommand{\OI}{[{O}{I}]}

\newcommand{\OIII}{[{O}{III}]}
\newcommand{\SII}{[{S}{II}]}
\newcommand{\NII}{[{N}{II}]}

\newcommand{\Ha}{H$\alpha\,$}
\newcommand{\Hb}{H$\beta\,$}

\newcommand{\NIIHa}{[{N}{II}]/H$\alpha$}
\newcommand{\SIIHa}{[{S}{II}]/H$\alpha$}
\newcommand{\OIHa}{[{O}{I}]/H$\alpha$}
\newcommand{\disp}{$\sigma$}

\shorttitle{}
\shortauthors{Rich et al.}

\begin{document}

\title{Galaxy-wide Shocks in Late-Merger Stage Luminous Infrared Galaxies}

\author{J. A. Rich \altaffilmark{1},  L. J. Kewley\altaffilmark{1} \&  M. A. Dopita\altaffilmark{2}\altaffilmark{1}}
\email{jrich@ifa.hawaii.edu}
\altaffiltext{1}{Institute for Astronomy, University of Hawaii, 2680 Woodlawn Drive, Honolulu, HI 96822}
\altaffiltext{2}{Research School of Astronomy and Astrophysics, Australian National University, Cotter Rd., Weston ACT 2611, Australia }

\date{\today}

\begin{abstract}
We present an integral field spectroscopic study of two nearby Luminous Infrared Galaxies (LIRGs) that exhibit evidence of widespread shock excitation induced by ongoing merger activity, IC 1623 and NGC 3256. We show the importance of carefully separating excitation due to shocks vs. excitation by HII regions and the usefulness of IFU data in interpreting the complex processes in LIRGs. Our analysis focuses primarily on the emission line gas which is extensive in both systems and is a result of the abundant ongoing star formation as well as widespread LINER-like excitation from shocks. We use emission-line ratio maps, line kinematics, line-ratio diagnostics and new models as methods for distinguishing and analyzing shocked gas in these systems. We discuss how our results inform the merger sequence associated with local U/LIRGs and the impact that widespread shock excitation has on the interpretation of emission-line spectra and derived quantities of both local and high-redshift galaxies.
\end{abstract}


\section{Introduction}

\begin{figure*}[htpb]
\centering
{\includegraphics[scale=0.95]{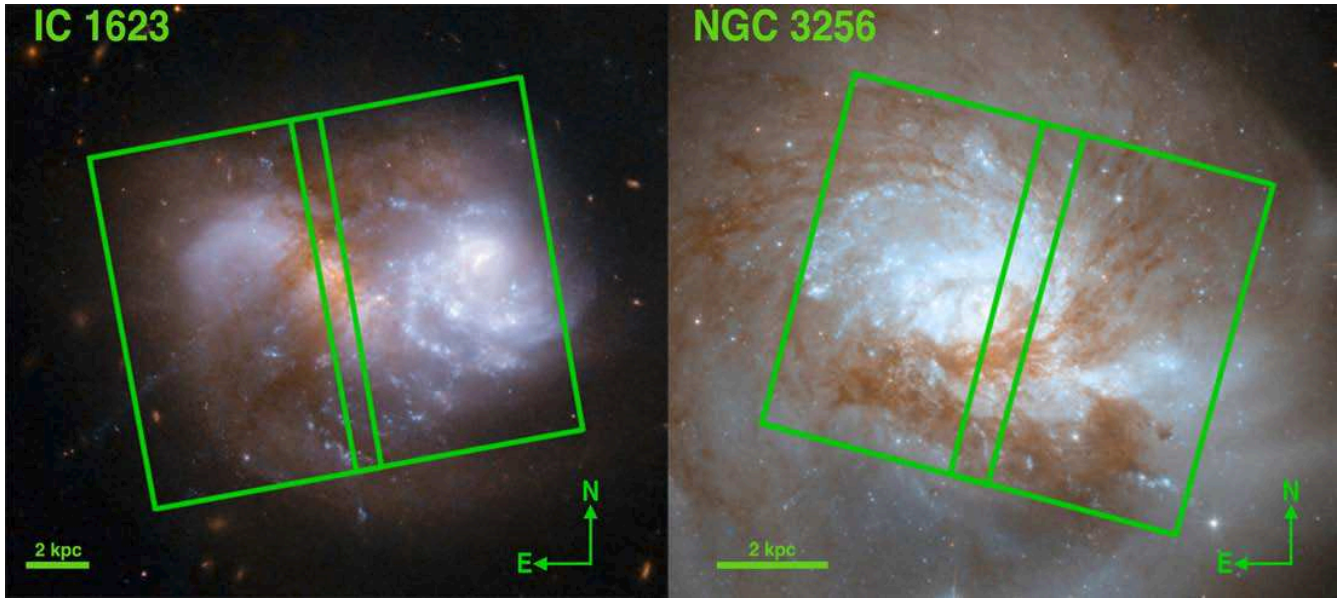}}
\caption{Archival HST ACS data of IC 1623 and NGC 3256 with our WiFeS pointings overlaid and 2 kpc bar for scale. The images shown are F435W + F814W, effectively Johnson B+I images. This filter combination distinguishes the intense star formation from the prominent dust lanes which in both galaxies heavily obscure the secondary nucleus of the merging pair.}
\label{fig1}
\end{figure*}

Understanding the physical processes involved in the formation and evolution of Galaxies remains a key problem in modern astronophysics. As observers strive to interpret data on the first generations of galaxies and theoreticians attempt to recreate the assembly and evolution of galaxies, studies of the nearby universe are needed to constrain key physical processes and global parameters that inform both observations and models of galaxies in the distant Universe. In this regard, Luminous Infrared Galaxies (LIRGs) and Ultraluminous Infrared Galaxies (ULIRGs) provide important observational constraints to models of distant galaxies. Although rare locally, IR-luminous galaxies and the merger activity associated with this class of systems become increasingly important at high redshift, contributing  \citep{Elbaz02,Lefloch05}. Not only does the improved spatial resolution allow us to study the relative importance of physical processes in the various components of the U/LIRG system, but also we can study the physics over the full array of merger states. Many recent studies across multiple wavelengths are beginning to focus on resolving the detailed ionization structure, kinematics and power sources of local U/LIRGs in order to further our understanding of galaxy evolution both locally and at high redshift.

The local population of LIRGs (which range from $10^{11}$ L$_{\odot}<$L$_{IR}<10^{12}$ L$_{\odot}$) have been found to represent a variety of objects from particularly dusty isolated star-forming galaxies to progressed mergers or late-tage mergers. Most frequently, however, U/LIRGs are often associated with merger activity which drives the intense circumnuclear star formation that fuels the powerful IR luminosity \citep{Veilleux95,Scoville00,Arribas04,Alonso06}. The gas in the outskirts of each system is torqued into the center which drives the ongoing star formation as the merger continues, \citep{Barnes96}. As the galaxies move along the \citet{Toomre72} merger sequence the IR luminosity is generally found to increase along with galactic wind activity \citep{Rupke05b}. At the terminal phase of the merger sequence galaxies have in general become ULIRGs, and appear as a coalesced or nearly coalesced system. Finally, these become dominated by a powerful AGN as their star formation becomes quenched as the available gas reservoir becomes depleted \citep{Genzel98,Yuan10}. 

As this merger sequence progresses, tidally induced gas motions and outflows from galactic winds become increasingly common. These forces can have an influence on the emission line gas through shocks that are induced by the large-scale gas flows \citep{Armus89,Heckman90,Lehnert96,Colina05,Zakamska10}. This shock excitation may contaminate line ratios used to determine metallicity, star formation rate and power source in a galaxy. Depending on the geometry and kinematics of the gas and dust the ionization of the observed emission line gas can be dominated by shock excitation induced by either tidal flows \citep{Colina05,Farage10,Monreal10} or galactic winds \citep{Veilleux02,Veilleux03,Lipari04,Sharp10,Rich10,Moiseev10}. Early work using narrow band or tunable filters and Fabry-Perot instruments proved useful for investigating these processes over large swaths of a galaxy \citep{Shopbell98,Veilleux02,Calzetti04}. However, a detailed characterization of the shock excitation and its effects on emission line spectra is more complex, necessitating the use of integral field spectroscopy (IFS).

The advent of integral field units (IFUs) has led to a rapid multiplication of observational evidence for extended shock excitation in nearby galaxies. \citet{Sharp10} conducted a study of several galactic winds and found widespread shock-excitation in the extended emission associated with the outflow. \citet{Monreal06} identified extended LINER-like emission resulting from shocks in a small IFU sample of ULIRGs and followed up with a more thorough IFU study of tidally induced shock excitation in nearby LIRGs \citep{Monreal10}. \citet{Farage10} discovered extended shocks caused by gas accreting onto a giant brightest cluster galaxy (BCG). Finally \citet{Rich10} found extended shock excitation caused by a galactic wind in the M82-like galaxy NGC 839. In both \citet{Farage10} and \citet{Rich10}, new slow shock models were employed to analyze the shocked gas. In all of the above cases shock excitation exhibits characteristics of extended LINER-like emission.

In this paper we present a detailed analysis of the widespread shock excitation in two galaxies drawn from a larger IFU sample of nearby U/LIRGs: the Wide Field Spectrograph (WiFeS) Great Observatory All-Sky LIRG Survey (GOALS) sample. We discuss these two systems, IC 1623 and NGC 3256 in \textsection~\ref{sample} and our observations, data reduction and line-fitting in \textsection~\ref{observations}.  We present the properties of the emission line gas for these two systems in \textsection~\ref{emissionlines}, including line ratio maps in \textsection~\ref{ratiomaps} and emission line diagnostic diagrams in \textsection\ref{diagnostics}. We discuss the distribution of velocity dispersions for each system in \textsection~\ref{dispersions}. We combine the observed quantities and use them to separate the shocked gas from the HII region gas and provide an analysis of our results in \textsection~\ref{analysis}. In \textsection~\ref{discussion} we discuss the implications of our results for the interpretation of emission line spectra and investigate the effect that extended shock excitation can have on quantities derived from single aperture spectra of nearby and high-redshift galaxies. Finally we give our conclusions in \textsection~\ref{conclusion}.

Throughout this paper we adopt the cosmological parameters H$_{0}$=70~km~s$^{-1}$Mpc$^{-1}$, $\Omega_{\mathrm{V}}$=0.72, and $\Omega_{\mathrm{M}}$=0.28, based on the five-year WMAP results \citet{Hinshaw09} and consistent with the \citet{armus09} summary of the GOALS sample. Systemic heliocentric velocities adjusted adjusted using the 3-attractor flow model of \citet{Mould00} are taken from \citet{armus09}.

\section{Sample}\label{sample}

The systems analyzed in this paper are drawn from a larger integral field spectroscopic survey of the GOALS sample. GOALS is a multi-wavelength survey of the brightest 60$\mu$m extragalactic sources in the local universe ($log(L_{IR}/L_{\sun}) > 11.0$), with redshifts z $<$ 0.088 and is a complete subset of the IRAS Revised Bright Galaxy Sample (RBGS) \citep{Sanders03}. Objects in GOALS cover the full range of nuclear spectral types and interaction stages and provide excellent nearby laboratories for the study of galaxy evolution.

\begin{deluxetable}{llllcc } 
\tablewidth{0pc}
\tablecolumns{6}
\tablenum{1} 
\tablehead{ \colhead{Galaxy}  &  \colhead{z}  & \colhead{R.A.} & \colhead{Decl.} & \colhead{B Mag.} &\colhead{log(L$_{IR}$/L$_{\odot}$)} }
\centering
\startdata
IC 1623      & 0.0201 & 01:07:47 & -17:30:27 & 15.04 & 11.71   \\
NGC 3256 & 0.00935 & 10:27:51 & -43:54:14 & 11.83 & 11.64   
\tablecaption{Information for the two systems observed for this paper. Names, systemic redshift and IR luminosities are taken from the \citet{Armus09} summary of the GOALS sample. RA and dec given are the center of our mosaiced WiFeS pointings. B magnitudes are from the surface photometry catalog of the ESO-Uppsala Galaxies.}
\enddata
\end{deluxetable} 

We chose IC 1623 and NGC 3256 for a more detailed analysis since they both show evidence of widespread shock excitation. Table 1 shows the global properties of the two systems, both of which are ongoing mergers of two massive spiral galaxies. Both systems are fairly luminous LIRGs and neither shows evidence of an embedded AGN. Archival HST data of the systems is shown in fig.~\ref{fig1}. Both systems show a complex optical morphology and contain one galaxy nucleus that is heavily extinguished by dust lanes. Both galaxies are classified as close mergers (projected separation between the merging nuclei of less than 10 kpc) when using the scheme of ~\citet{Veilleux02}, which utilizes classifications based on numerical simulations of mergers ~\citep{Barnes92,Barnes96}.

\textbf{IC 1623:}
IC 1623 (also commonly referred to as VV 114 and with the IRAS designation IRASF 01053-1746) consists of two galaxies: IC 1623 A in the west and the more heavily obscured IC 1623 B in the east. The dust lane seen in figure~\ref{fig1} covers the eastern nucleus, which is actually brighter in the near-IR than the western disk \citep{knop94,Doyon95,Scoville00}. The galaxy nuclei are separated by approximately 5 kpc, firmly classifying the pair as a close merger in the ~\citet{Veilleux02} classification scheme.  Interestingly, the heavily obscured eastern galaxy has two very bright sources of similar magnitude in the near-IR, raising the possibility that IC 1623 B itself is a merger. However, it may be that there is simply a second region of heavily obscured star formation unassociated with the true nucleus. Because of the extensive dust lane, the majority of the optical emission in our observations is dominated by the star formation in the western galaxy. Neither of the galaxies exhibits evidence of an AGN, obscured or otherwise ~\citep{Doyon95,Veilleux95}. Extended tidal features can be seen in optical and near-IR imaging, though the low surface brightness of these features makes observations of them prohibitively time-consuming. The galaxy pair is at a redshift of z$\sim 0.0201$, corresponding to a 1\arcsec projected distance of $\sim$385 pc. 

\textbf{NGC 3256:}
NGC 3256 (IRAS designation IRASF10257-4339) is a late-stage merger with nuclei separated by approximately 1 kpc, though it has not yet formed a single nucleus and therefore cannot be classified as a post-merger. \citet{Sakamoto06} note that the long tidal tails seen in optical and HI images \citep{deVaucouleurs61,English03b} suggest ``a prograde-prograde merger of two gas-rich spiral galaxies of similar size''  \citep{Toomre72,White79}. The bright face-on galaxy seen in figure~\ref{fig1} has a separated, heavily obscured secondary nucleus seen in x-ray, IR and radio lying $\sim$5\arcsec\ south of the bright optical nucleus \citep{Boeker97,Neff03,Sakamoto06}. \citet{Neff03} posited that the two nuclei may be low-luminosity AGN based on the radio to x-ray ratios, though they also note that a large number of supernova remnants could also account for the observed ratio. However subsequent observations by \citet{Lipari04b} of galactic bubbles associated with previous supernovae in the core of NGC 3256 support the latter case and rule out any significant contribution from even a low-luminosity AGN. This supernova activity in the core of NGC 3256 is also driving a galactic wind measured in neutral absorption and ionized emission previously associated with shock heating \citep{Armus89,Heckman90,Heckman00,Lipari00}. NGC 3256 lies closer than IC 1623 at z$\sim 0.0094$, which means that our data probe a finer spatial scale of 1\arcsec $\sim185$ pc.

\section{Observations \& Data Reduction}\label{observations}

Our data were taken using the Wide Field Spectrograph (WiFeS) at the Mount Stromlo and Siding Spring Observatory 2.3 m telescope.  WiFeS is a dual beam, image-slicing integral field spectrograph commissioned in May 2009 and described in detail by \citet{Dopita07} and \citet{Dopita10}. Our data consists of separate blue and red spectra with wavelength coverage of $\sim$3500-5800 \AA~ \& $\sim$5500-7000 \AA~respectively and a resolution of R = 3000 (blue) and R = 7000 (red), corresponding to a velocity resolution of 100\kms at \Hb and 40\kms at \Ha. Observations of IC 1623 were carried out on August 20, 2009. Two separate pointings were observed for 40 minutes in total for each pointing. NGC 3256 was observed on March 15, 2010 with two individual pointings for 60 minutes each. Each galaxy was observed using two pointings  as shown in ~\ref{fig1}, with a large enough overlap region to facilitate mosaicing. The effective field of view of the data in both cases is thus approximately $45\arcsec \times 38\arcsec$, providing well over 1,000 spectra in each set of data cubes.

The data were reduced and flux calibrated using the WiFeS pipeline, briefly described in \citet{Dopita10}, which uses IRAF routines adapted primarily from the Gemini NIFS data reduction package. A single WiFeS observation consists of 25 slit spectra (and an additional 25 contemporaneous sky spectra if the data was taken in nod and shuffle mode). Each individual observation was reduced into a data cube using the process described below. 

Bias subtraction is slightly complicated due to temporal changes in the quad-readout bias levels and a slope in the bias level on each WiFeS chip. Bias frames are taken immediately before or after each observation or set of observations and a 2-D fit of the surface of this bias frame is subtracted from the temporally nearest object data in order to avoid adding additional noise to the data. Any resulting residual in the bias level is accounted for with a fit to unexposed regions of the detector.

Quartz lamp flats are used to remove pixel to pixel variations from the response curve of the chip and twilight sky flats are used to correct for illumination variation along each of the slitlets.  Spatial calibration is carried out by placing a thin wire in the filter wheel and illuminating the slitlet array with a continuum lamp. This procedure defines the center of each slitlet. The individual spectra have no spatial distortion because the camera corrects the small amount of distortion introduced by the spectrograph. Thus only low-order mapping of the slitlets is required. 

CuAr and NeAr arc lamps are used to wavelength calibrate the blue and red spectra respectively. Arc lamp data were taken throughout each night of observing to account for any change in the wavelength solution. Each of the 25 slitlets is then rectified by the pipeline into a full data cube (one for each arm) and sampled to a common wavelength scale for each target. The spectral resolution achieved within the red data cubes is $46\pm6 \kms$. We use this value to remove instrumental broadening from the derived velocity dispersions.

Telluric absorption features were removed from the resulting red data cubes using observations of B-stars or featureless white dwarfs taken at similar air mass. The effects of differential atmospheric refraction are calculated and corrected by the pipeline for the blue data cubes.

Flux calibration of each individual data cube was carried out using flux standards observed throughout each night. The white dwarf EG 131 was used to flux calibrate IC 1623 and the white dwarf L745-46a was used to flux calibrate NGC 3256 \citep{Bessell99}.

Finally, the individual reduced, flux-calibrated data cubes are median combined and sampled to a common spatial grid using overlapping features found in each pointing. Observations of NGC 3256 were  binned on-chip by 2 pixels in the spatial direction in order to increase signal to noise and produce square spatial elements $1\arcsec \times 1$\arcsec. Observations of IC 1623 were binned after reduction for the same reasons. This means that for both systems the scale of the data cubes and of the resulting maps presented in this paper are $1\arcsec$/spaxel.

Combined data cubes were aligned astrometrically by comparing a pseudo 'r-band' image generated using the red spectrum from each spaxel with HST images from the archive.

\subsection{Spectral Fitting}

\begin{figure*}[h]
\centering
{\includegraphics[scale=0.5]{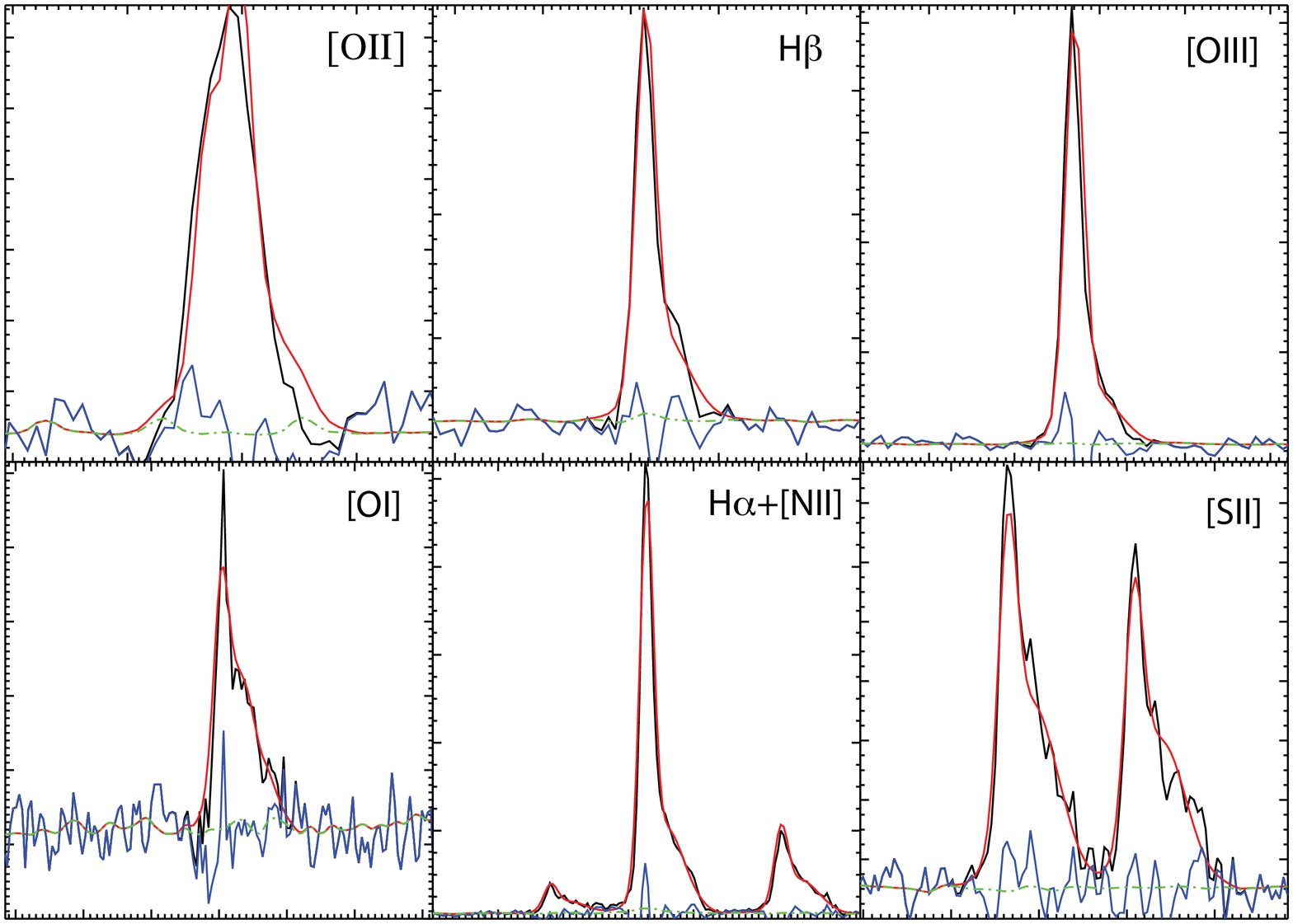}
 \includegraphics[scale=0.5]{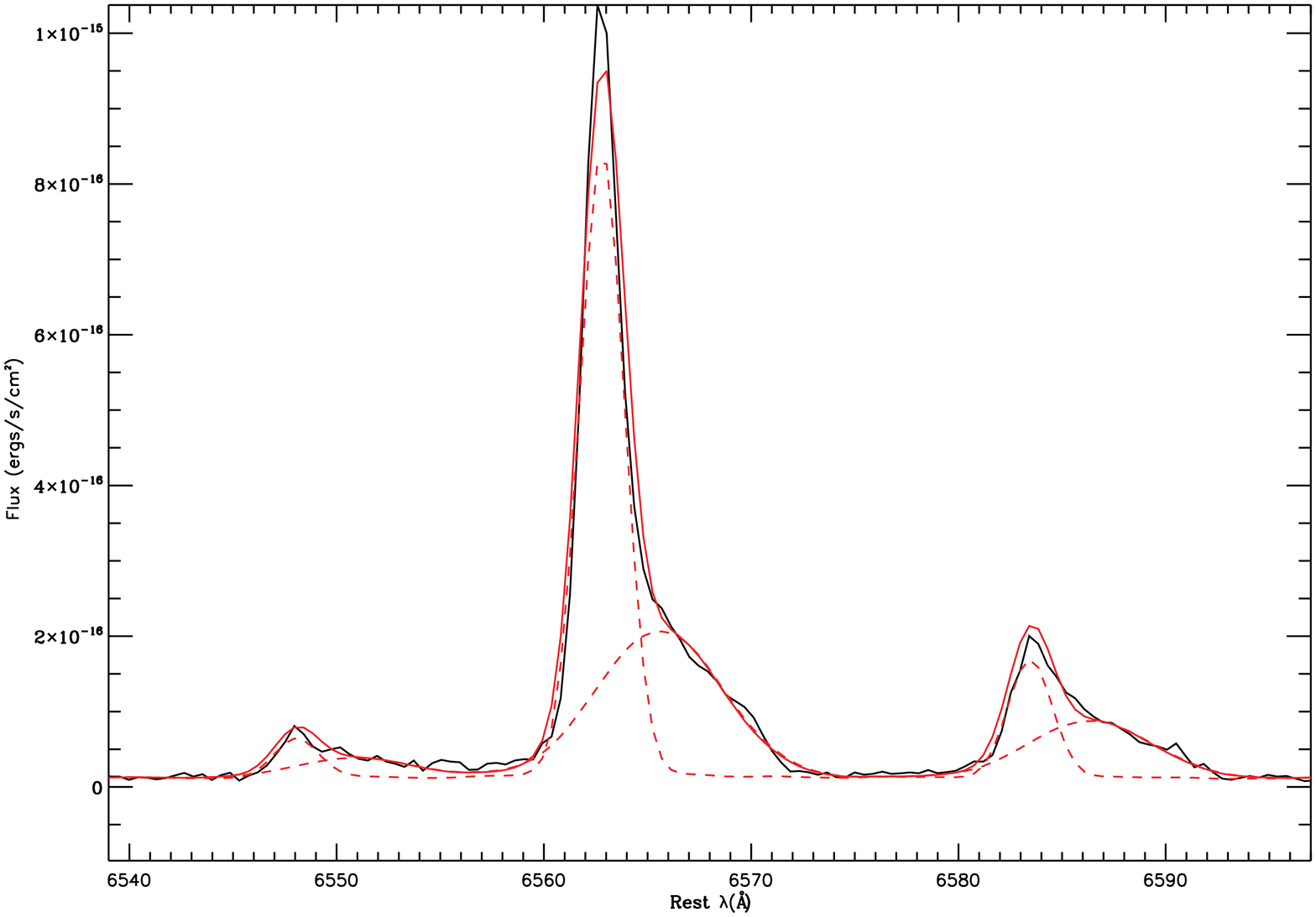}}
\caption{Example of emission line fitting routine results for a single spectrum from IC 1623.  The upper panels show individual emission lines of interest with data in black, total fit in red, continuum in green, and residuals in blue. The bottom panel shows a zoomed in region around \NII~and \Ha with each Gaussian component plotted as a dotted line and the total fit plotted as a solid line. Wavelengths are plotted in the rest frame with respect to the narrow component.}
\label{fig2}
\end{figure*}

We analyzed every spectrum using an automated fitting routine written in IDL, UHSPECFIT, which is based on the code created by \citet{Zahid10} and is also employed by and described in ~\citet{Rupke10b} and ~\citet{Rich10}. Our routine fits and subtracts a stellar continuum from each spectrum using population synthesis models from \citet{Gonzalez05} and an IDL routine which fits a linear combination of stellar templates to a galaxy spectrum using the method of \citet{Moustakas06}. 

After the continuum is removed, lines in the resulting emission spectra are fit using a one or two-component Gaussian, depending on the goodness of fit determined by the routine. Resulting fits were inspected to ensure the fitting code did not fail. All emission lines are fit simultaneously using the same Gaussian component or components. Both continuum and emission lines were fit using the MPFIT package, which performs a least-squares analysis using the Levenberg-Marquardt algorithm \citep{Markwardt09}. Example fits are provided in fig.~\ref{fig2}.

The errors in the parameters used in fitting the emission lines are calculated by the fitting code. This includes the gaussian widths (velocity dispersions) and ratio of dispersions between the blue and red spectra, individual gaussian peaks, redshift and any slight deviation between the blue and red wavelength calibrations.  Emission line fluxes used in derived quantities are subject to a cut of minimum signal-to-noise of 5, which is calculated by comparing the peak of the Gaussian to the noise in the continuum. The lower resolution of the blue spectra contributes to the uncertainty of the gaussian parameters-for instance in some cases 1 or 2 emission line components for a blue emission line may provide a similar goodness of fit. By minimizing the $\chi^{2}$ for all of the lines simultaneously we increase slightly the overall uncertainty in the velocity dispersion while maintaining the ability to analyze multiple components fits.

In many cases it is crucial that more than one Gaussian is used in the emission line fit. Our analysis relies on a careful decomposition of multiple velocity profiles, which are common in the kinematically complex family of U/LIRGs. An example can be seen in fig.~\ref{fig2}, where two distinct components have been fit. In the case shown, there is an obvious red-shifted component with a broader profile than the primary component. The line ratios between the two components are noticeably different as well.

\section{Emission Line Gas Properties}\label{emissionlines}
\begin{figure*}[htpb]
\centering
{\includegraphics[scale=0.82]{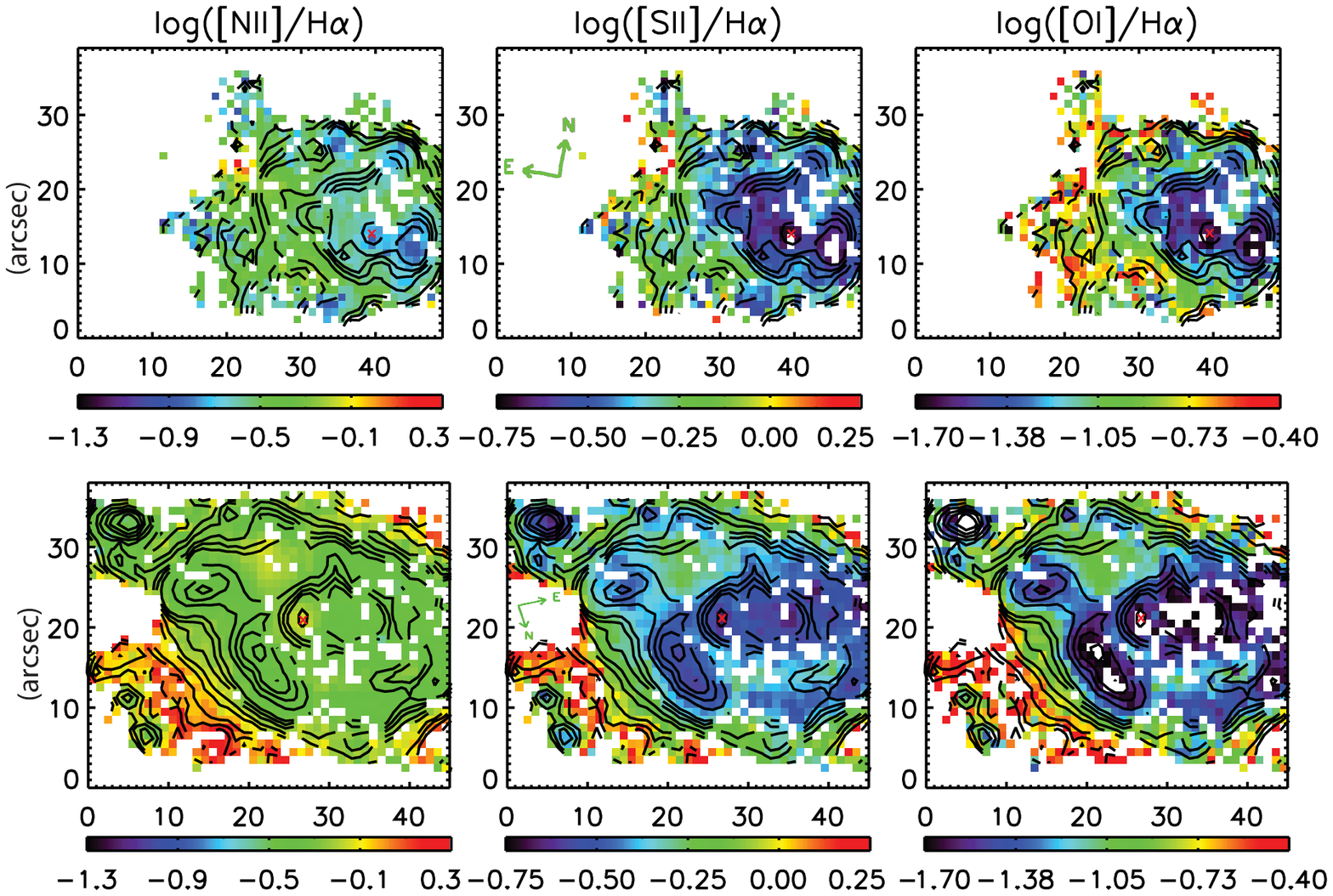}}
\caption{From left to right, emission line ratio maps of \NII/\Ha, \SII/\Ha and \OI/\Ha for IC 1623 (top) and NGC 3256 (bottom). \Ha contours have been overlaid, and the peak in \Ha flux has been marked with a small red "x". These maps are a sum of the total emission line flux in both profiles and trace the dominant ionization mechanism in different portions of the galaxy. Generally speaking, the stronger the line ratio the further the departure from an ionizing spectrum purely due to star formation. The regions most dominated by ongoing star formation appear bluer in the above maps. Maps are oriented with respect to the original observations (cf. fig. 9)}
\label{fig3}
\end{figure*}

Emission line ratios have proven a useful tool for probing the sources of ionizing radiation in galaxies. The primary factors influencing emission line strengths are the shape of the extreme ultraviolet (EUV) spectrum and the abundance of the emission line gas \citep{Kewley01b,Dopita00}. Thus with IFU data, a line ratio maps and diagnostic diagrams can be created as a useful tool in interpreting the power sources not just in the nucleus of a galaxy but also across the galaxy as a whole.

\begin{figure*}[htpb]
\centering
{\includegraphics[scale=0.85]{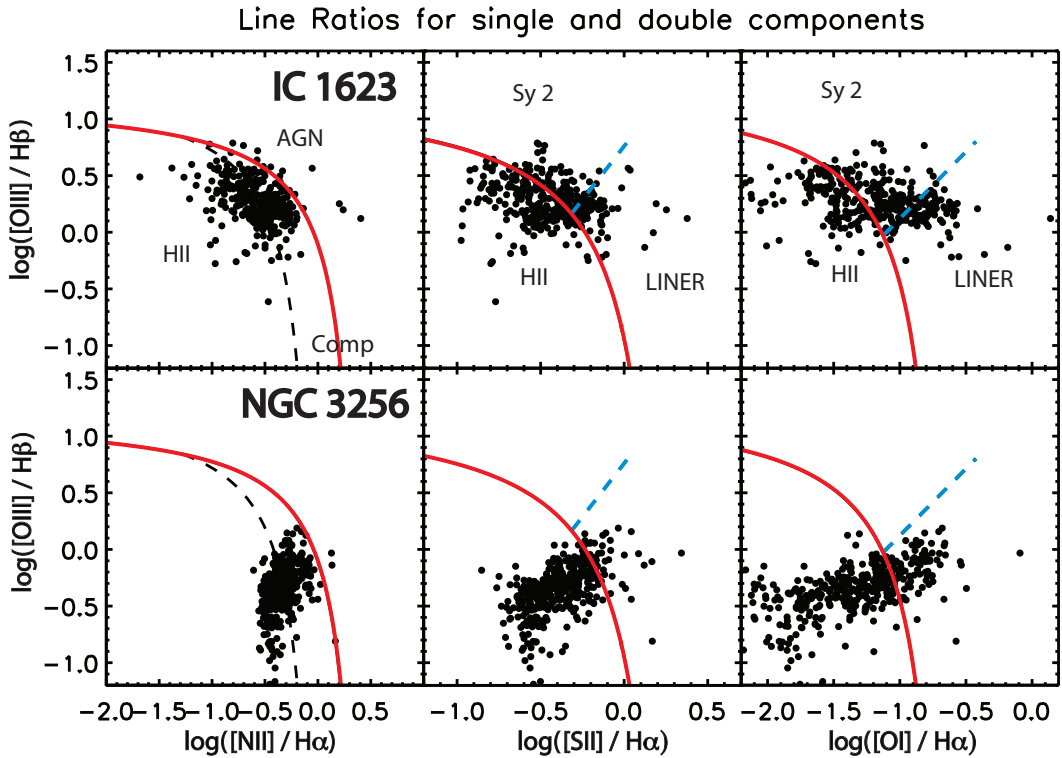}}
\caption{Emission line ratio diagnostic diagrams of each individual fitted emission line component for each system. Boundaries are taken from \citet{Kewley06}. We apply a minimum S/N cut of 5 in every diagnostic line. The resulting points in each diagram reflect a mix of ionization by HII region emission and by shock excitation. The points lying in the HII region portion of the diagram also reflect the abundance of the gas: the HII region gas in NGC 3256 is higher metallicity resulting in a different overall shape to the distribution of points in the diagnostic diagram.}
\label{fig4}
\end{figure*}

\subsection{Line Ratio Maps}\label{ratiomaps}
In figure 3, we examine maps of the ratios of the total flux of \NII\ 6583 \AA, \SII\ 6717,6731\AA, and \OI\ 6300\AA\ to \Ha\ which are sensitive to metallicity as well as the ionization parameter as a first step towards understanding the processes at work in IC 1623 and NGC 3256. 

\textbf{IC 1623:}
Most of the optical emission line flux in IC 1623 is dominated by the eastern galaxy (IC 1623 A), which is less extinguished by dust. The emission line ratios in this portion of the merging system are lower than in the rest of the system, tracing the star formation evidenced by the less extinguished HII regions and clusters seen in the archival HST images. There is no evidence of a metallicity gradient in the \NII/\Ha ratio map, which is in agreement with recent work on the flattening of metallicity gradients in merging systems by ~\citet{Rupke10a,Rupke10b,Kewley10}.

Moving outward from the main regions of unobscured star formation in IC 1623 A we see an overall rise in all line ratios. Towards the northern and southern parts of IC1623 A, the line ratios increase to levels inconsistent with excitation by HII regions alone until the signal to noise in emission is too low. Especially striking are the high line ratios correlated with the dust lane east of IC 1623 A which obscures the better part of IC 1623 B. The emission line ratios show consistently high values across this dust lane, especially in \OI/\Ha, which is a particularly good tracer of shock excitation ~\citep{Farage10,Rich10}. The total \OI\ flux in this region approaches 1/4 to 1/3 the total \Ha flux, while the total \NII~flux is as much as 1/2 to 2/3 the total \Ha flux.

\begin{figure}[h]
\flushleft
{\includegraphics[scale=0.27]{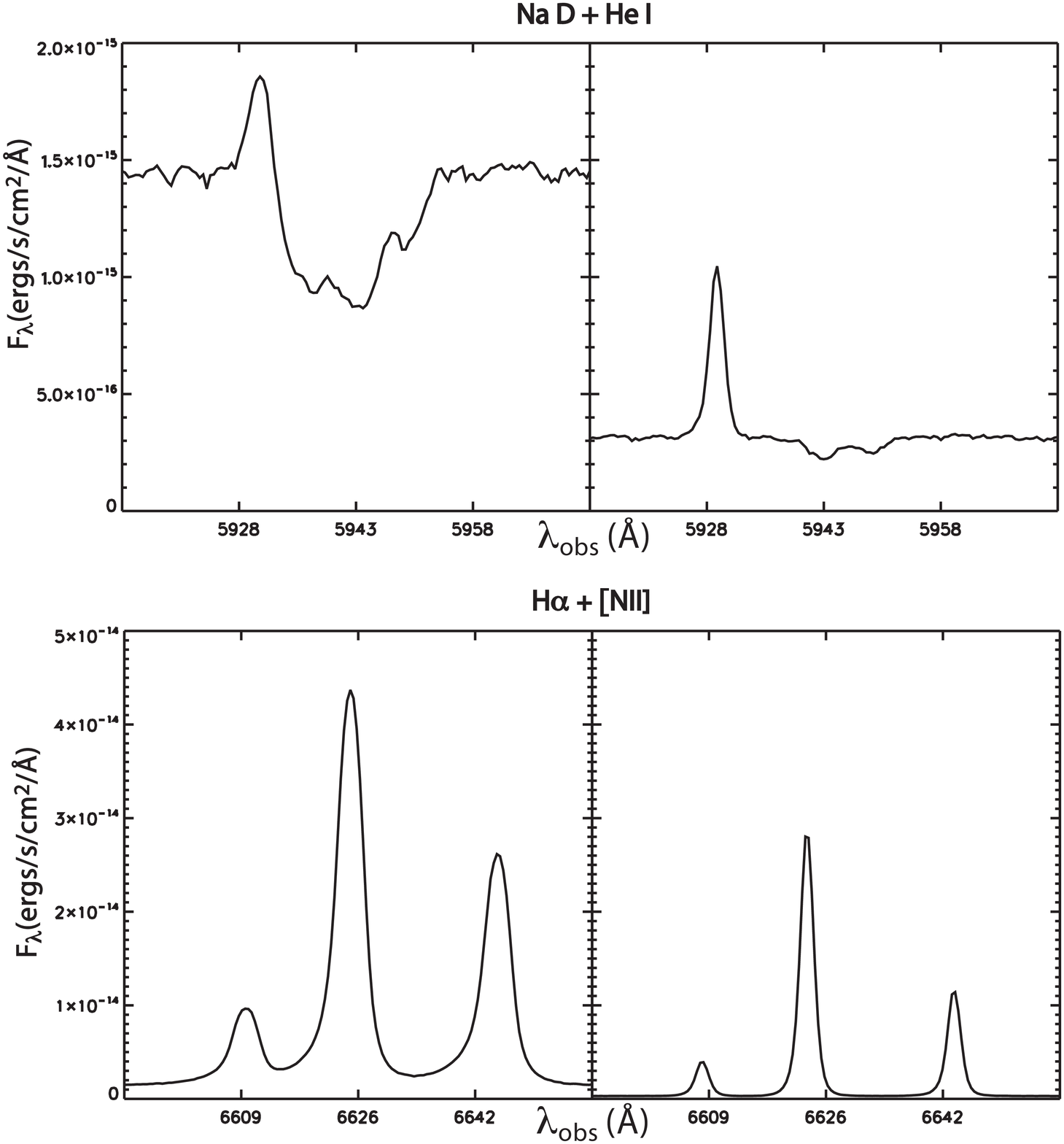}}
\caption{Line profiles from 3" extracted apertures from NGC 3256 for the brightest optical nucleus (left) and an off-nuclear region dominated by star formation (right). The nuclear spectrum shows a large blue-shifted Na D component as well as a corresponding blue tail in emission both caused by the large galactic wind driven by the nuclear starburst.}
\label{fig5}
\end{figure}

\textbf{NGC 3256:}
The line ratio maps of NGC 3256 show similar behavior compared to IC 1623. Again the areas of unobscured star formation correspond to lower line ratios, tracing the ongoing star formation as the dominant source of ionizing photons. A key difference is that NGC 3256 shows a region of increasing line ratios in a small central part of the region of low line ratios. This increase corresponds to the previously observed galactic wind in NGC 3256 ~\citep{Heckman90} and our spectra in this region exhibit strong, broad blue-shifted NaD profiles (fig. \ref{fig5}). The increasing emission line ratios in this case are due to the additional presence of shock excitation from the galactic wind.

Outside of both the central region influenced by the galactic wind and the primarily star formation dominated portions of the galaxy the line ratios again tend to increase, with \NII/\Ha reaching nearly unity and correspondingly high \SII/\Ha and \OI/\Ha in the outer spatial portions of our data cubes. These regions of strong excitation correlate strongly with dust lanes evident in the HST archival data and are quite distant from the core of the galaxy.

\subsection{Diagnostic diagrams}\label{diagnostics}
In order to better understand the power sources at play, we turn to line ratio diagnostic diagrams. Diagnostics using \NII/ \Ha,\SII/\Ha or \OI/\Ha ratios against the  \OIII/\Hb ratio were first employed by \citet{BPT81} and \citet{VO87} to distinguish the source of ionizing photons in global or nuclear galaxy spectra. This diagnostic scheme has been successfully applied to Integral field spectra (as well as narrow-band and tunable filter data) to investigate emission line gas over entire galaxies or regions of a galaxy to probe the distribution of ionizing sources (e.g. \citealt{Shopbell98,Sharp10,Rich10}). To separate the ionizing sources with IFU data, individual spectra from each spaxel are plotted on the traditional diagnostic diagrams. In this way the ensemble of points provides information about the various processes at work in different portions of the galaxy rather than a single global measure.

We plot diagnostic diagrams in figure~\ref{fig4} and include all emission line components from each spaxel with S/N$>5$ in all 6 diagnostic lines (\Hb, \Ha, \OIII\ 5007\AA, \OI\ 6300\AA, \NII\ 6583\AA, \SII\ 6717,6731\AA). In all diagrams, the lower left-hand portion of the plot traces photionization by HII regions.  The solid curved line traces the upper theoretical limit to pure HII region contribution measured by ~\citet{Kewley01b} and the dotted line in the \NII~diagnostic provides an empirical upper limit to the HII region sequence of SDSS galaxies measured by ~\citet{Kauffmann03}. The region lying between these two lines represents objects with a composite spectrum mixing HII region emission with a stronger ionizing source. Contribution from a Seyfert AGN pushes points into the upper region of the diagrams while more LINER-like emission lies to the right hand side of the diagrams, as shown by the dividing line in the \SII\ and \OI\ diagnostics, as established in \citet{Kewley06}. New lower velocity shock models of order 100-200 \kms\ trace line ratios from the upper edge of HII-like emission into the LINER region of the diagnostic diagrams \citep{Farage10,Rich10}.

\section{Velocity Dispersions}\label{dispersions}
Where more than one component is fit to emission lines ordering the components over a whole IFU data cube is non-trivial. The difference in velocity dispersion and flux where two components are fit varies depending on the conditions in the region of the galaxy observed and is dependent on several factors including the relative flux, source of ionization and extinction within each component. In some cases two components of comparable flux and velocity dispersion are observed, while in other regions two components of significantly different flux, flux ratios and velocity dispersion are seen (cf \ref{fig2}). In this section we focus on the velocity dispersion of individual emission line components as the most useful way of probing the multiple components observed

If shock excitation is important in these regions showing enhanced \OIHa, \NIIHa and \SIIHa ratios, then we would expect to see the presence of these shocks reflected in increased velocity dispersions in these areas. As the emission line ratios increase in a manner consistent with shocked gas, the velocity dispersion should increase to velocities consistent with the speed of the shocks, which are of order 100-200 \kms. Lower emission line ratios consistent with gas in HII regions should exhibit lower velocity dispersions, typically of a few 10s of \kms \citep{Epinat10}. In a similar study of LIRGs with IFUs, \citet{Monreal06, Monreal10} observed the correlation between excitation and velocity dispersion consistent with widespread shock excitation in a larger sample of nearby U/LIRGs and associated this excitation with tidally induced shocks.

\begin{figure}[h]
\centering
{\includegraphics[scale=0.65]{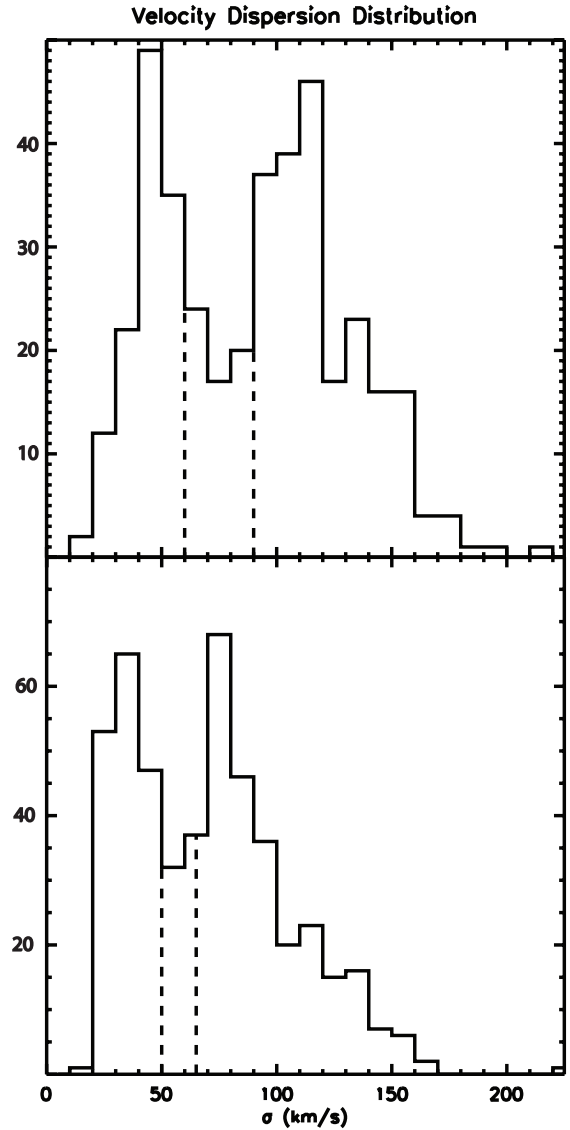}}
\caption{Distribution of emission line velocity dispersions from every individual Gaussian component fit. The bimodal distribution of velocity dispersions between the low velocity (unresolved HII region) and high velocity (shock-excited) components can be seen quite clearly and the cutoffs between low and high velocity dispersion are marked with dotted lines. The bimodal distribution in NGC 3256 is less sharply defined due to the combination of outflow and inflow.}
\label{fig6}
\end{figure}

\begin{figure*}[htpb!]
\centering
{\includegraphics[scale=0.85]{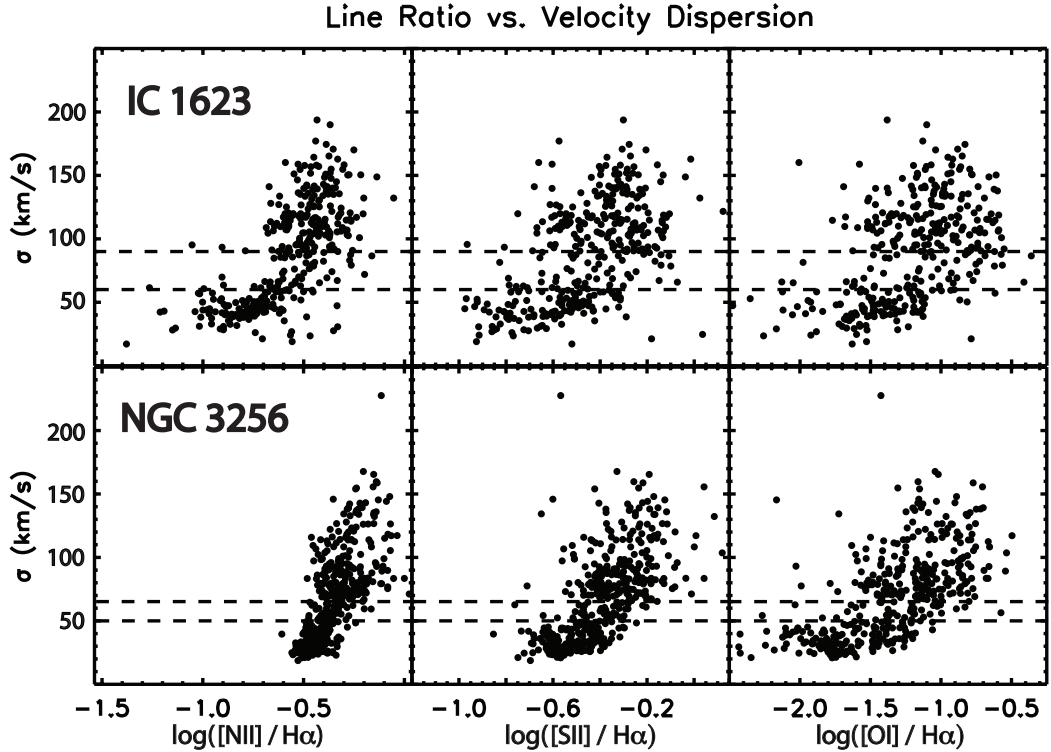}}
\caption{Plot of line ratio vs. velocity dispersion for each individual emission line component.  Regions of emission line gas dominated by shock excitation show both higher emission line ratios and higher velocity dispersion resulting in the correlation seen above. A sequence of narrow component, low-excitation points corresponds to HII region gas of varying ionization parameters and metallicities. There is less scatter in the \NIIHa~line ratio owing to the higher signal to noise of that line, though the broader spread of \SIIHa~and especially \OIHa~makes these ratios a better tracer of shock excitation. }
\label{fig7}
\end{figure*}

We are able to more accurately separate the lower velocity dispersion (low-\disp) emission consistent with HII regions and the higher velocity dispersion (high-\disp) with our multiple component emission line fitting, as large portions of the gas in both systems exhibit a composite of activity both in emission line ratios and in velocity dispersion. For instance, in IC 1623 the gas falling on to the dust lane is well fit by a single fairly broad component, while gas elsewhere in the galaxy is mostly fit by either two narrow components or a narrow and broad component consistent with the complex velocity field and multiple-phase gas in this merger.

In fig.~\ref{fig6} we plot a histogram of velocity dispersions for each system for every component fit by our routine. The spectrum from every spaxel was fit with either one or two Gaussian profiles with unique widths, and all of those components are plotted here. We only plot values from spaxels with emission lines of S/N$>5$ in all of our diagnostic lines,  \Hb, \OIII, \OI, \NII, \SII\, and \Ha. We have removed the effect of instrumental broadening from our velocity dispersions.

In both systems there is a clear peak at low-\disp~($\sim40 \kms$ consistent with a large contribution from HII region emission due to the substantial amount of ongoing star formation. In the case of IC 1623 there is a clear bimodality between the low-\disp~(HII region) and high-\disp~(shocked) components. The bimodality in NGC 3256 is still present but appears less sharply delineated owing to a combination of shock excitation by the galactic wind and by bulk gas motions due to the ongoing merger.

We use the histograms in fig.~\ref{fig6} to establish a cutoff between the low-\disp~and the high-\disp~gas. For each system dashed lines show the cuts in fig.~\ref{fig6} which separate the emission line components into low-\disp, transition and high-\disp~to better trace the different power sources. For IC 1623, we consider all gas with \disp~$< 60$ \kms~as low-\disp~and all gas with \disp\ $> 90$ \kms as high-\disp~with the transition falling in between these two values.  Similarly, we establish cutoffs of low-\disp~at~$<50$\kms and high-\disp~at~$>65$\kms for NGC 3256.

Figure~\ref{fig7} shows velocity dispersion vs. emission line ratio for each individual component. Correlation between these two measures of shock excitation was explored using single-component fits to IFU data of several LIRGs by \citet{Monreal10}. The increasing velocity dispersion correlated with increasing emission line ratio corresponds to the dominance of shock excitation at high-\disp, while the narrow sequence of low-\disp~points (bottoming out at the instrumental resolution) corresponds to emission from HII region gas. 

\section{Analysis}\label{analysis}

\begin{figure*}[htpb]
\centering
{\includegraphics[scale=0.85]{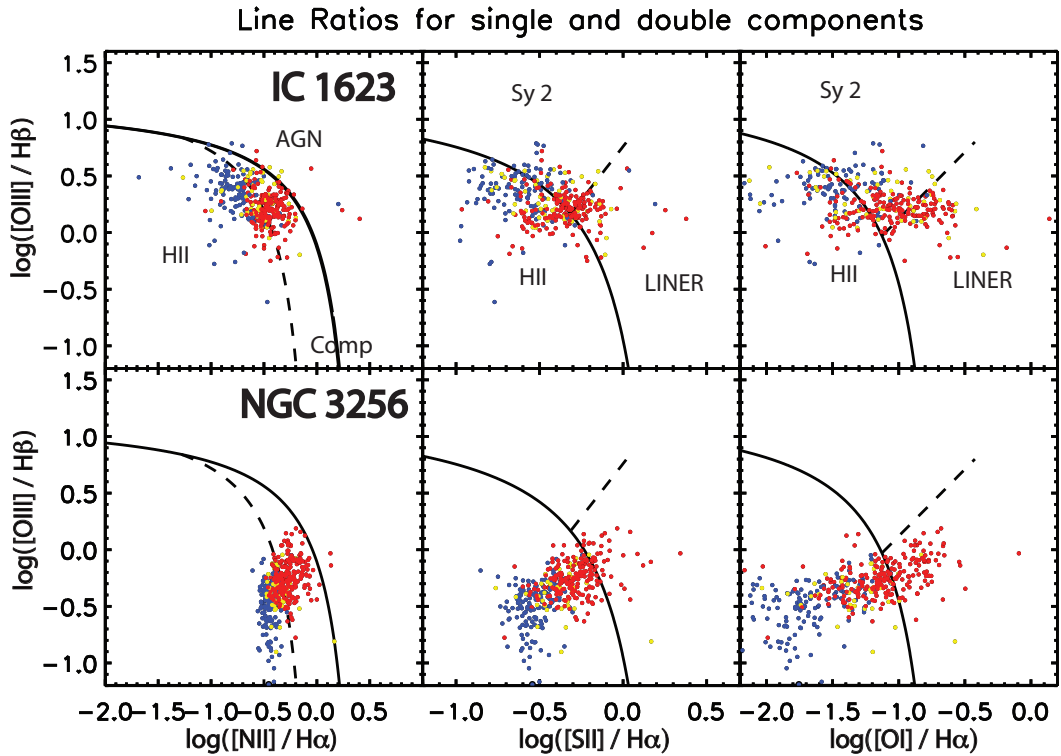}}
\caption{Same plot as fig.~\ref{fig4}, but with points color coded based on velocity dispersion cutoffs established in section~\ref{dispersions}. Low-\disp\ points are plotted in blue, transition points in yellow and high-\disp\ points in red. Clearly the low velocity (HII region) and the high velocity (shocked) components are characterized by different ranges of line ratios.}
\label{fig8}
\end{figure*}

In figure~\ref{fig8} we color-code the points in our diagnostic diagrams from figure~\ref{fig4} using the low-\disp\ and high-\disp\ cutoffs for each system in order to separate the HII region gas from the shocked gas. The correlation of velocity dispersion with line ratio strength is still clearly seen, most obviously in the \NII~diagnostic owing to the higher S/N of that line. More importantly, there is a clear sequence from HII-region to LINER in the \SII and \OI diagrams with a corresponding change from low-\disp\ to high-\disp\ gas tracing the influence of shocks. The locus of the HII region emission-line gas points in the \NII~diagnostic diagram begins in the middle/low metallicity region for IC 1623 and in the higher metallicity region for NGC 3256, corresponding to the abundances of the HII regions in those galaxies.

Color coding the points accentuates the mixing sequence from pure HII region emission to pure shock excitation already seen in the emission line diagnostic diagrams. By combining emission line diagnostic ratios and velocity dispersion analysis of individual emission line components we can perform a fairly conservative separation of the HII region gas and the shocked gas of each system and analyze them separately.

\begin{figure*}[htpb]
\centering
{\includegraphics[scale=0.9]{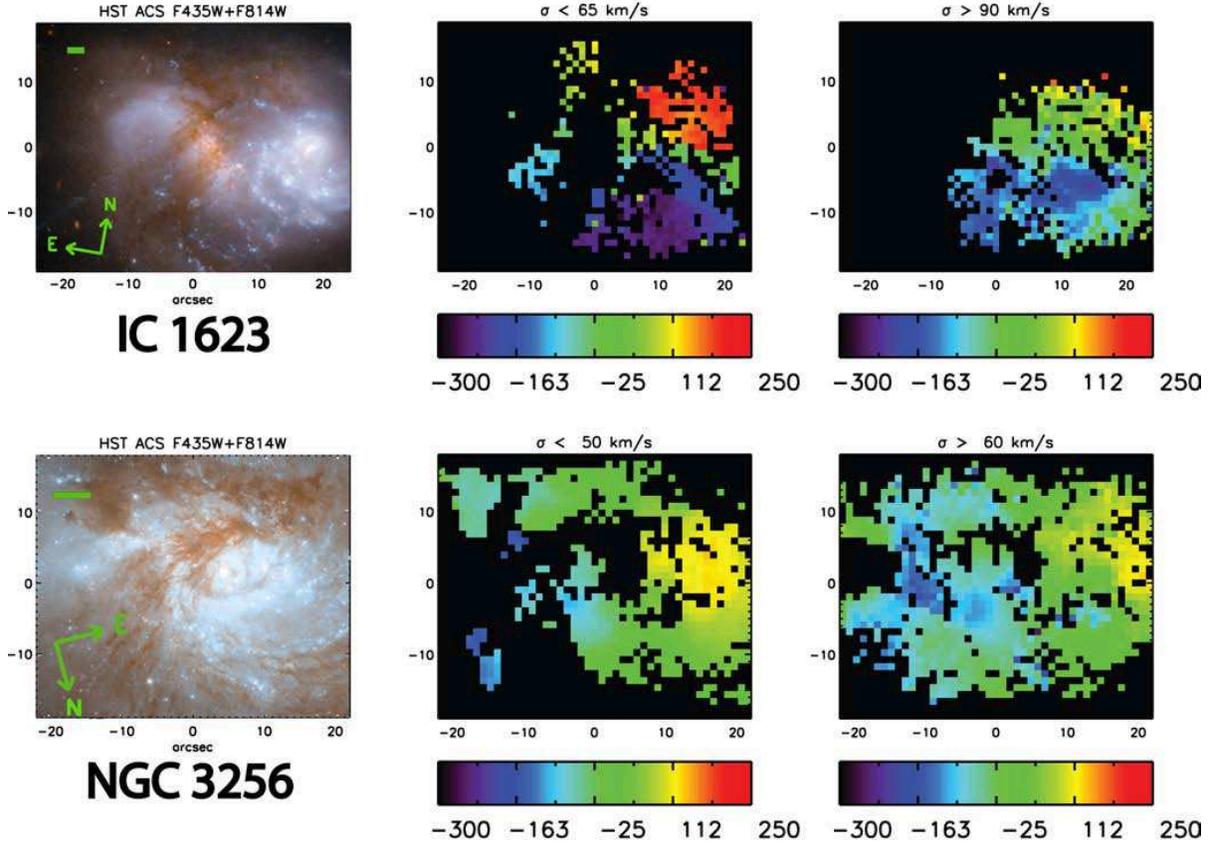}}
\caption{The above figure shows the line-of-sight velocities in \kms of the HII region (low-\disp) gas and the shocked (high-\disp) gas for IC 1623 (top) and NGC 3256 (bottom). The leftmost panel is the HST ACS image from fig.~\ref{fig1} for reference. The HII region gas in both systems traces the rotational motion of the galaxy. The shocked gas in IC 1623 shows a dramatically different kinematic profile associated with the bulk motions of the shocked gas with components both blue and red shifted from the HII region component. Overall the broad component in NGC 3256 is slightly blue shifted with respect to the HII region gas in the star forming regions and slightly red shifted along the dust lanes.}
\label{fig9}
\end{figure*}

In fig.~\ref{fig9} we plot the line of sight velocity of the low-\disp\ and the high-\disp\ separately for both systems. The first thing of note is the presence of a high-\disp\ component over nearly the entirety of both systems-clearly shocks are widespread in these galaxies. We have excluded some points from the central regions of each system where our fitting routine was unable to satisfactorily fit 2 Gaussians to the complex line profiles or where both emission line components are broad enough to exceed the shock cutoff. In both systems the low-\disp\ gas traces the HII regions which follow the rotational motion of the parent galaxies. 

In IC 1623 the low-\disp\ gas is primarily dominated by the rotation of the eastern galaxy and is comparable to the overall shape of the rotation observed in the CO gas observed by \citet{Yun94}. There is a gap in low-\disp\ gas consistent with the dust lane seen in the HST images, as the shock excitation is correlated with the dust lane. In this region  our emission-line analysis only traces the tidally-induced shock excited gas between the two systems because the underlying star formation of the western galaxy is completely extinguished by dust. The kinematics of the shocked gas are quite distinct, exhibiting very little correlation with the rotation seen in the low-\disp\ gas. The shocked gas in the northern portion of IC 1623a is blue shifted up to 200

The kinematic distinction between the HII region gas and the shocked gas is more subtle in NGC 3256. Where both a shocked component and an HII region component are both measured, however, the shocked gas is mostly blue-shifted. Over the larger part of the visibly star forming regions of NGC 3256 there is a mild shocked component blue shifted a few 10s of \kms. The dust lane extinguishes the HII region gas as in IC 1623 and in spaxels where both a low-\disp\ and a high-\disp\ component are measured along the dust lane the shocked gas is generally redshifted a few 10s of \kms. In addition, NGC 3256 hosts a galactic wind seen quite clearly in blue shifted Na D \citep{Heckman90,Heckman00}. The spectra affected are within a few arcseconds of the nucleus and have a very broad blueshifted tail extending several hundred \kms and a broader narrow component such that both components have strong line ratios and are considered high-\disp\ and thus excluded from the velocity plots. Finally, there is a region $\sim8$\arcsec in diameter of high-\disp\ component strongly blue shifted-up to nearly 200 \kms-lying northeast of the nucleus in a region with little dust and several closely-packed star clusters, possibly a second localized outflow.

\section{Discussion}\label{discussion}
Our analysis clearly separates the HII region gas from the shocked gas where both are detected. In both systems along the dust lanes and in the periphery of the galaxies where star formation either decreases or is extinguished the emission line spectra are dominated by a single, shocked component. However, our results indicate that even in the regions where emission is dominated by star formation there is still an underlying broad component associated with shocks, induced either by outflows generated by the circumnuclear star formation or by the tidal motions of the gas. 

\subsection{Shock and HII region Models}
To characterize this shock excitation we employ new slow shock models which reproduce our measured emission-line ratios. These models are introduced in \citet{Farage10} and \citet{Rich10} and will be discussed in detail in an upcoming paper (Kewley et al. in prep). In short we generate slow shock models with velocities consistent with our observed line widths using an updated version of the Mappings III code, originally introduced in \citet{Sutherland93}. In addition, we compare HII region models generated using Starburst99 \citep{leitherer99} and Mappings III. Both sets of models have consistent abundances of solar and twice solar for IC 1623 and NGC 3256 respectively, using the abundance set of \citet{grevesse10}. HII models are plotted with varying ionization parameter $Q(H)$, the flux of ionizing photons divided by the hydrogen atomic density (the dimensionless ionization parameter $U = Q/c$), while shock models are shown at varying shock velocities.

\begin{figure*}[htpb]
\centering
{\includegraphics[scale=0.38]{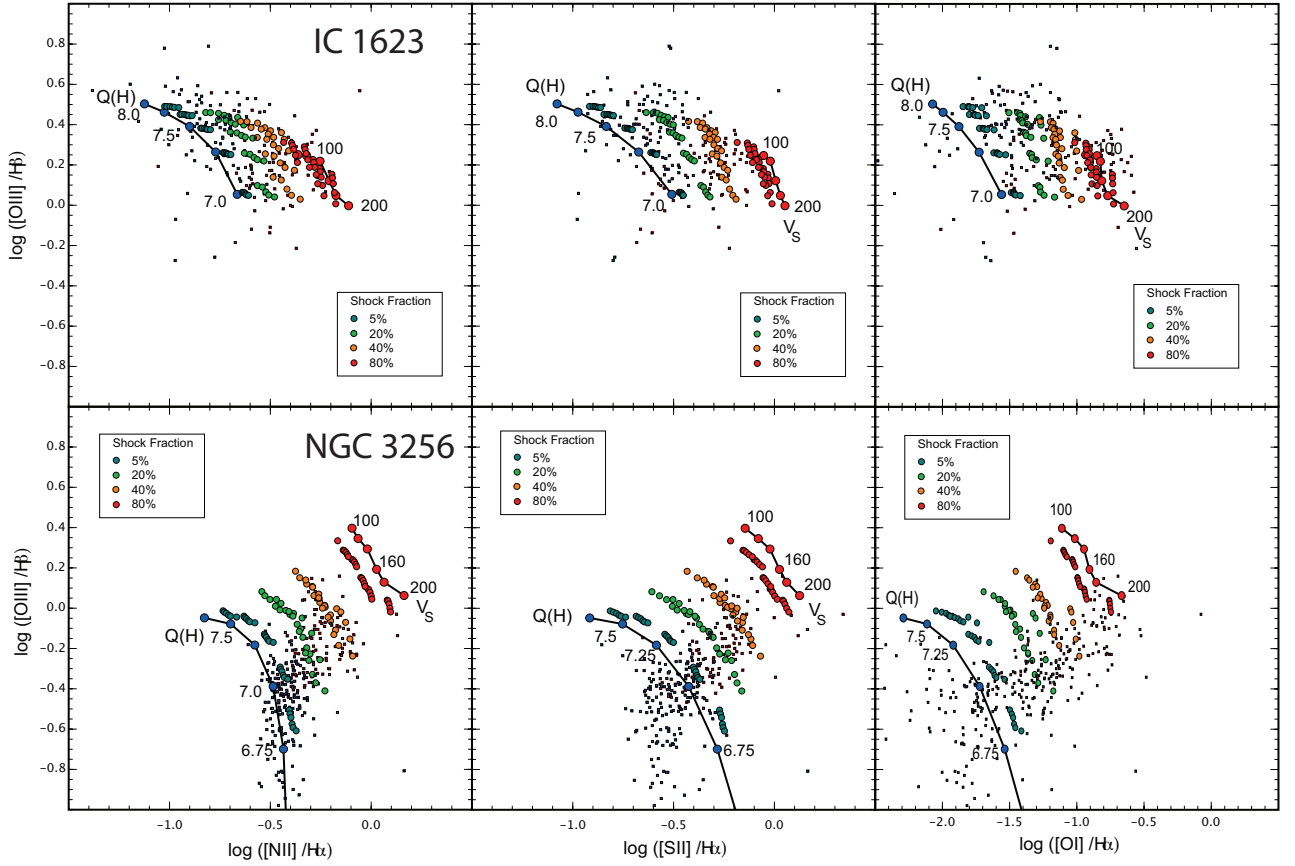}}
\caption{Here we replot measured emission line ratios (small points) with our HII region and shock models overlaid. The HII region models are larger blue circles connected by a black line and are plotted with varying ionization parameter (Q) and adopting abundances of solar and twice solar respectively for IC 1623 and NGC 3256. The fully preionized slow shock models are plotted as larger red circles and are also connected by a black line along a sequence of increasing shock velocity.  The colored points in between show a mixing sequence from pure HII region to pure shock excitation.}
\label{fig10}
\end{figure*}

These models are plotted in figure~\ref{fig10} with the measured emission line ratios underplotted as small points. As in figure~\ref{fig8}, the points lying furthest to the left are more consistent with pure star formation. Our HII region models reproduce these line ratios. The data for IC1623 is consistent with a range in ionization parameters between $7.0<log[Q(H)] < 8.0$, while the higher-abundance NGC 3256 is characterized by a lower mean ionization parameter $6.75 <log[Q(H)] < 7.25$ consistent with the dynamical models of HII regions in different metallicity environments presented by \citet{SEDII,SEDIII}. As shocks become a more dominant ionizing mechanism and/or as the light from HII regions is extinguished by dust allowing shocks to dominate the line ratios migrate along the theoretical mixing sequence shown in fig.~\ref{fig10}. 

The shape of the mixing sequence evident in figs.~\ref{fig8}~\&~\ref{fig10} within the diagnostic diagrams is caused by the different underlying abundances in the two systems. In IC 1623 the lower-abundance gas means as shock excitation begins to dominate the line ratios move horizontally across the diagrams to the LINER region. The relatively higher abundance of NGC 3256 however creates a tilted sequence with the pure HII region points beginning in a different region of the diagnostic diagrams. This difference is most noticeable in the \NII~ diagnostic diagram where the different metallicities are most widely separated in line ratio space.

Varying the fractional contribution of shock excitation within the parameters of our models recovers the observed emission line ratios quite well. It is important to note in~\ref{fig10} that the line ratios are sensitive to as little as 20\% shock-excitation. The observed ratios in IC 1623 cover nearly the entire mixing sequence, implying that the various regions sampled in the galaxy cover a range of shock velocities, though most of the points near the end of the mixing sequence correspond to 100 to 140 \kms.  In NGC 3256 the line ratios both within the mixing sequence and near the pure shock models are consistent with shock velocities of $\sim160-200$ \kms. In both systems the terminus of our fastest shock models shows remarkable agreement with the strongest measured emission line ratios.

\subsection{Global Energetic Considerations}
It is worthwhile to make a few simple calculations to quantify the impact that shocks have on dissipating the energy of the infalling gas in merging systems. Here we focus on the well-studied NGC 3256, where 
previous wide-scale observations facilitate these calculations. 

Once the shocked emission in our data has been isolated we can use the measured \Ha flux provided solely by shocks to extrapolate the total shock luminosity. For this calculation we make the simple assumption that the total $F_{H_{\alpha}}$ measured in any line components above the shock velocity dispersion threshhold is caused by shocks.  We use the plot provided in~\citet{Rich10} to approximate the total shock luminosity from $L_{H_{\alpha}}$ originating in shocks. This provides an upper limit to the total shock luminosity of 80x L$_{H_{\alpha}}$. For NGC 3256 this corresponds to an approximate global shock luminosity of $L_{shock}\sim 8\times10^{42} ergs~s^{-1}$

Working with this luminosity as a starting point we next consider the total gas available as well as the mass and size of the system. ~\citet{English03a, English03b} report an I-band (stellar) mass within 70" of the system of~$(2.5\pm0.5)\times10^{10}h^{-1}M_{\odot}$ and a total HI gas mass of~$3.5\times10^{9}h^{-2}M_{\odot}$, 75\% of which resides in tails extending to a total distance of $42h^{-1}$kpc. This does not take into account the molecular gas mass, which may be larger. Adopting our value of h=0.7 provides a dynamical timescale of $\sim800 Myr$ and a total binding energy for the HI gas of $\sim8\times10^{56}ergs$.

The star formation rate can be approximated using L$_{H_{\alpha}}$ and L$_{IR}$ \citep{Kennicutt98,SEDI,Dopita03,Calzetti07,Kennicutt09}. ~\citet{Lira02} estimate the total F$_{H_{\alpha}}\sim2\times10^{-11}ergs~s^{-1}~cm^{-2}$, with data from ~\citet{Lipari00} using a larger aperture than is available to our IFS data. Applying a total reddening correction of $E(B-V)=0.5$, approximated using maps of the balmer decrement from our data cubes, yields L$_{H_{\alpha}} \sim1.2\times10^{43}ergs~s^{-1}$.   Assuming a fractional contribution to \Ha from shocks of $\sim30\%$, consistent with our IFS aperture, the total star formation rates are $\sim66 M_{\odot}/yr$ and $\sim75 M_{\odot}/yr$ using L$_{H_{\alpha}}$ and L$_{IR}$~respectively. The addition of both extinction correction and an adjustment due to shock contribution yields a better agreement in the star formation rates. If a SFR of 75 $M_{\odot}/yr$, persists the available HI gas can fuel the starburst at this rate for $\sim100 Myr$.

For shocks to aid in removing kinetic energy from the infalling gas, we consider the amount of time it would take for the shock luminosity to dissipate about twice the binding energy of the available HI gas, or $\sim10^{56}ergs$. If the entirety of the the shock luminosity of $L_{shock}\sim 8\times10^{42} ergs~s^{-1}$ acts in this way, the kinetic energy dissipation time scale is $\sim6$ Myr, significantly less than the dynamical time scale, implying that shocks may act as a very efficient mechanism for removing kinetic energy from the infalling gas, allowing it to feed the central starburst. \citet{English03a}, however, do not consider inclination effects in their mass calculations, which would effectively increase this timescale.

\subsection{Star Formation Rate}
The \Ha flux is often used as a direct proxy for star formation rate in galaxies, typically assuming the relation of \citet{Kennicutt98}. If one assumes that all of the high-\disp\ shocked emission in each system is coming from pure shock excitation, a simple total fractional contribution from shocks and thus the effect on the measure of star formation rate can be estimated.

Internal extinction will also impact the measurement of \Ha-derived SFR and may affect regions of shock excitation differently than regions dominated by HII photoionization. When comparing the balmer decrements from our data for both IC 1623 and NGC 3256, the average difference between these regions is negligible. In some portions of the galaxy where both high and low \disp components are seen the extinction in the starburst component is slightly stronger than in the shocked component, though the average difference in $E(B-V)$ is comparable to the intrinsic spread in values.

After applying the velocity dispersion cutoffs, the total \Ha flux of the high-\disp\ emission line gas accounts for as much as 1/3 of the total \Ha emission, which would decrease an estimated star formation rate by $\sim30\%$. Clearly shocked emission must be considered when SFR measurements are made from \Ha in galaxies with starburst driven winds and merger widespread shock excitation

\subsection{Aperture \& Redshift}
As shocks contribute fractionally to the emission line fluxes over the majority of the galaxy and dominate the total flux in some off-nuclear regions, care must be taken when deriving quantities from the emission line spectra. In order to better understand the effect of aperture size and it's changing physical scale with redshift, we artificially redshifted our data cubes while extracting a psuedo-fiber spectrum from the peak in median rest-frame red light flux. 

\begin{figure*}[htpb]
\centering
{\includegraphics[scale=1.00]{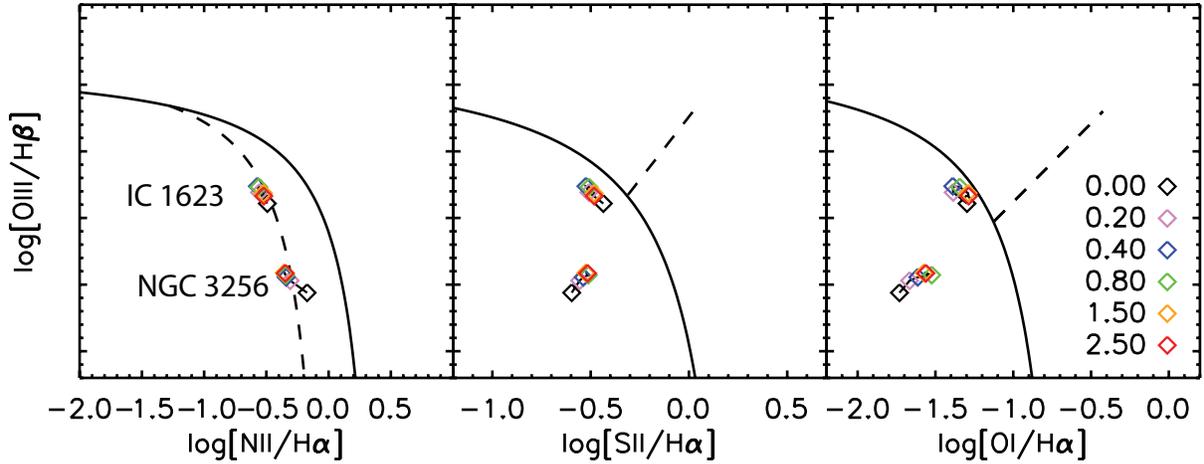}}
\caption{Single component fit to artificially redshifted, SDSS fiber aperture-extracted data cubes. Both IC 1623 and NGC 3256 are shown and points are color coded according to redshift. The points do not shift more than $\sim0.4$ dex. NGC 3256 moves from a composite classification to an HII classification in the \NII diagnostic.}
\label{fig11}
\end{figure*}

We choose SDSS as our point of comparison, using a 3\arcsec~fiber size for our aperture extractions and rebinning our spectra in wavelength to match the approximate R$\sim1800$ spectroscopic resolution of SDSS. We maintained a spaxel size of 1\arcsec while spatially rebinning and dimming the data cubes to correspond to z of 0.2, 0.4, 0.8, 1.5 and 2.5. After extracting a composite pseudo-fiber spectrum at each redshift, we apply our fitting routine using both one and two component Gaussian fits to emission lines. The measured emission line ratios for each redshift and each component are shown in figure \ref{fig11}. Asymmetries in the emission lines are in some cases fit somewhat better by 2 components, but the overall separation between the two components is $\sim100 \kms$, within the simulated resolution.

As we redshift our data, the emission line ratios do not change by more than 0.4 dex at the most. Most of this shift occurs after the first redshift owing to the binning of a large part of the galaxies emission. Both galaxies move from a more composite to a more starburst-like classification in the \NII diagram. The two component fit is more difficult to interpret, but generally the broader component in both systems moves to the right in the diagnostic diagrams (more shock like) while the narrower component moves to the left (more HII region like). Because of the low separation in velocity space the interpretation of the secondary component is probably unreliable.

With such an SDSS fiber aperture at the highest simulated redshifts (z=1.5, 2.5) the spectra become a sum of nearly all of the emission line gas in the system and the total emission line fluxes of the weaker lines decrease to a few $\times10^{-17}$ergs s$^{-1}$ cm$^{-2}$ or less, making them marginally detectable. Proportionally more of the high-\disp\ gas is of lower surface brightness, so that while more of the outer regions dominated by high-\disp\ shocked gas are coadded, their total surface brightness drops, countering this. The total velocity dispersions of the narrow portion of the emission lines also becomes broader as more HII region ensembles are coadded, increasing to higher-\disp\ of $\sim100$ consistent with those measured in high-z galaxies. This effect was noted using IFU observations by \citet{Epinat10} and causes difficulty in discerning between large dispersion clumps and effects and the coadding of several low-\disp\ regions at high redshift. This is especially noticeable in IC 1623 which has a complex velocity structure across the majority of the brightest portions of the galaxy. 

Because of these effects, separating the shocked gas component from the star forming component within global spectra is difficult at the highest z. Improvement can be achieved with higher spatial and spectral sampling coupled with deep observations, though the purely shocked high-\disp\ regions associated with the dust lanes in NGC 3256 and IC 1623 are of lower surface brightness making them difficult to detect even if the global spectra show composite behavior. 

\section{Conclusion}\label{conclusion}

Using IFU emission line data of two nearby LIRGs, we have found a bimodal distribution of velocity dispersions and emission line ratios indicative of composite activity. This composite emission can be explained by a combination of shocks and star formation. In order to separate the shocked component from the purely star forming component we establish cutoffs in velocity dispersion to distinguish the shocked higher-\disp\ component from the HII region-like lower-\disp\ gas. Once this distinction has been made, the HII region like gas can be used to calculate derived quantities based on HII region emission line data such as star formation rate and metallicity. The properties and overall contribution of the shocked gas can also be inferred from the emission line data as well. We have provided such an example using low-velocity shock models which accurately reproduce the observed emission line ratios at the extreme end of the emission line ratio diagnostics. The shape of the mixing sequences between these models is directly influenced by the gas abundance and our HII-region models imply a distribution of Q(H) consistent with with the dynamical models for HII regions by \citet{SEDII,SEDIII}.

We urge caution when interpreting emission line data of both nearby and high-z galaxies, especially when spatial information is limited. Traditionally the classification of a galaxy relies on a nuclear spectrum, or a globally extracted aperture in the case of more distant systems. Using this method our systems would be classified as composite locally and starburst at higher-z, though this classification is affected by the size and position of the aperture extracted and the morphology of the system. Slit spectra provide a slightly more complete picture though the alignment and size of the slit and the size and distance of the system may complicate interpretation. This is especially true at high redshift (z$\sim1.5-3$) where a slight misalignment of the slit coupled with the drastic reduction in surface brightness of most features makes such measurements difficult and may miss the complete picture.

Our data clearly illustrates the importance of shocks in merging galaxies-not only do they affect quantities derived from spectroscopy, they also act as a useful mechanism for dissipating the kinetic energy and angular momentum of the infalling gas in merging systems. This aids in driving the gas to the center of the galaxy in order fuel the ongoing central starburst and keeping it in an increasingly deep potential well and presumably feed the eventual AGN. Presumably a portion of the shock excitation is generated by galactic winds as the merger progresses, though even a few percent of the total shock luminosity we calculate is sufficient to dissipate the binding energy on a time scale shorter than the dynamical timescale (10 to 100 Myr vs. 800 Myr).  As the gas is continually driven inwards and shocked by tidal forces, galactic outflows coupled with star formation will also compete to use up the gas on the scale of a few hundred million years or so.

Our work also illustrates the utility of Integral Field Spectroscopy in accurately decomposing the complex aspects of a merger. Not only is the full spatial picture of many properties available with an IFU observation, it is easier to separate the widespread shocked component and to trace its spatial and kinematic structure. Once this information has been isolated, the derived quantities of the HII region emission can be more accurately calculated and the detailed properties of the shocked gas itself can be investigated. This is especially important when calculating metallicities from strong-line calibrations, a quantity of particular interest in merging galaxies locally and in the early universe. A forthcoming paper (Kewley et al. in prep) will provide a theoretical framework for the effect of shock emission on such measurements and discuss the mixing sequence in galaxies such as those presented in this paper. We also plan to apply this careful separation of shocked gas from HII region emission where necessary to the measurement of metallicities in a larger sample of IFU data of merging galaxies in a separate paper (Rich et al. in prep).

\begin{acknowledgements}
We thank the referee for his/her helpful comments, which helped clarify several points in this paper. Dopita acknowledges the support from the Australian Department of Science and Education (DEST) Systemic Infrastructure Initiative grant and from an Australian Research Council (ARC) Large Equipment Infrastructure Fund (LIEF) grant LE0775546 which together made possible the construction of the WiFeS instrument. Dopita would also like to thank the Australian Research Council (ARC) for support under Discovery  project DP0664434. Dopita, Kewley and Rich acknowledge ARC support under Discovery  project DP0984657. This research has made use of the NASA/IPAC Extragalactic Database (NED) which is operated by  the Jet Propulsion Laboratory, California Institute of Technology, under contract with the National  Aeronautics and Space Administration.  This research has also made use of NASA's Astrophysics Data System, and of SAOImage DS9 \citep{joye03}, developed by the Smithsonian Astrophysical Observatory.
\end{acknowledgements}


\bibliographystyle{apj}

\end{document}